\documentclass[aps,prl,twocolumn,groupedaddress,showpacs]{revtex4-1}
\usepackage{graphicx}  % needed for figures
\usepackage{dcolumn}   % needed for some tables
\usepackage{bm}        % for math
\usepackage{amssymb,amsfonts,amsmath}   % for math
\usepackage{color}
\usepackage{lipsum}

\usepackage[normalem]{ulem}

\renewcommand{\d}{\mathrm{d}}

\newcommand{\bs}{\boldsymbol}
\renewcommand{\d}{\mathrm{d}}
\renewcommand{\v}{{\boldsymbol v}}

\newcommand{\w}{{\boldsymbol \omega}}
\allowdisplaybreaks

\begin{document}

%\title{The Random Walk of Choanoflagellate Colonies}
%\title{The Random Walks of Multicellular Choanoflagellates}
\title{Motility of Colonial Choanoflagellates\\ and the Statistics of Aggregate Random Walkers}
%\title{Multicellular Choanoflagellates are Ideal Composite Random Walkers}
%\input author_list.tex       % D0 authors (remove the first 3 lines
                             % of this file prior to submission, they
                             % contain a time stamp for the authorlist)
                             % (includes institutions and visitors)

\author{Julius~B.~Kirkegaard}
\author{Alan~O.~Marron}
\author{Raymond~E.~Goldstein}
\affiliation{Department of Applied Mathematics and Theoretical Physics, Centre for Mathematical Sciences, \\ University of 
Cambridge, Wilberforce Road, Cambridge CB3 0WA, United Kingdom}

\date{\today}


\begin{abstract}
We illuminate the nature of the three-dimensional random walks of microorganisms 
composed of individual organisms adhered together.
Such \textit{aggregate random walkers} are typified by
choanoflagellates, eukaryotes that are the closest living relatives of animals.
In the colony-forming species {\it Salpingoeca rosetta} we show 
that the beating of each flagellum is stochastic and uncorrelated with others, and 
the vectorial sum of the flagellar propulsion manifests as stochastic 
helical swimming. A quantitative theory for these results is presented and species variability discussed.
\end{abstract}

\maketitle

Active microparticles, self-propelled by stored energy or that available from the environment, 
typically exhibit directed motility combined with rotational diffusion, leading to random walks 
that at large times are statistically similar to their equilibrium counterparts. 
For artificial swimmers 
such as Janus particles \cite{Walther2013}, powered by inhomogeneous surface chemical reactions, the source of 
randomness is 
the same thermal 
fluctuations that translate Brownian particles, but here rotate them \cite{Howse2007}. In biology, 
several paradigms for stochastic locomotion exist. For single-celled organisms, 
stochastic beating leads to noisy swimming paths \cite{Ma2014}, and
active processes such as flagellar 
bundling/unbundling by bacteria \cite{Berg1993} and synchronization/desynchronization in algae \cite{Polin2009} enhances this stochasticity.  `Obligate' eukaryotic polyflagellates  
such as the ciliate \textit{Paramecium} \cite{Michelin2010} and the 
alga \textit{Volvox carteri} \cite{Brumley2012}, exhibit large-scale flagellar coordination, 
and increased regularity of motion.    

\begin{figure}[b]
\centering
\includegraphics[width=0.48\textwidth]{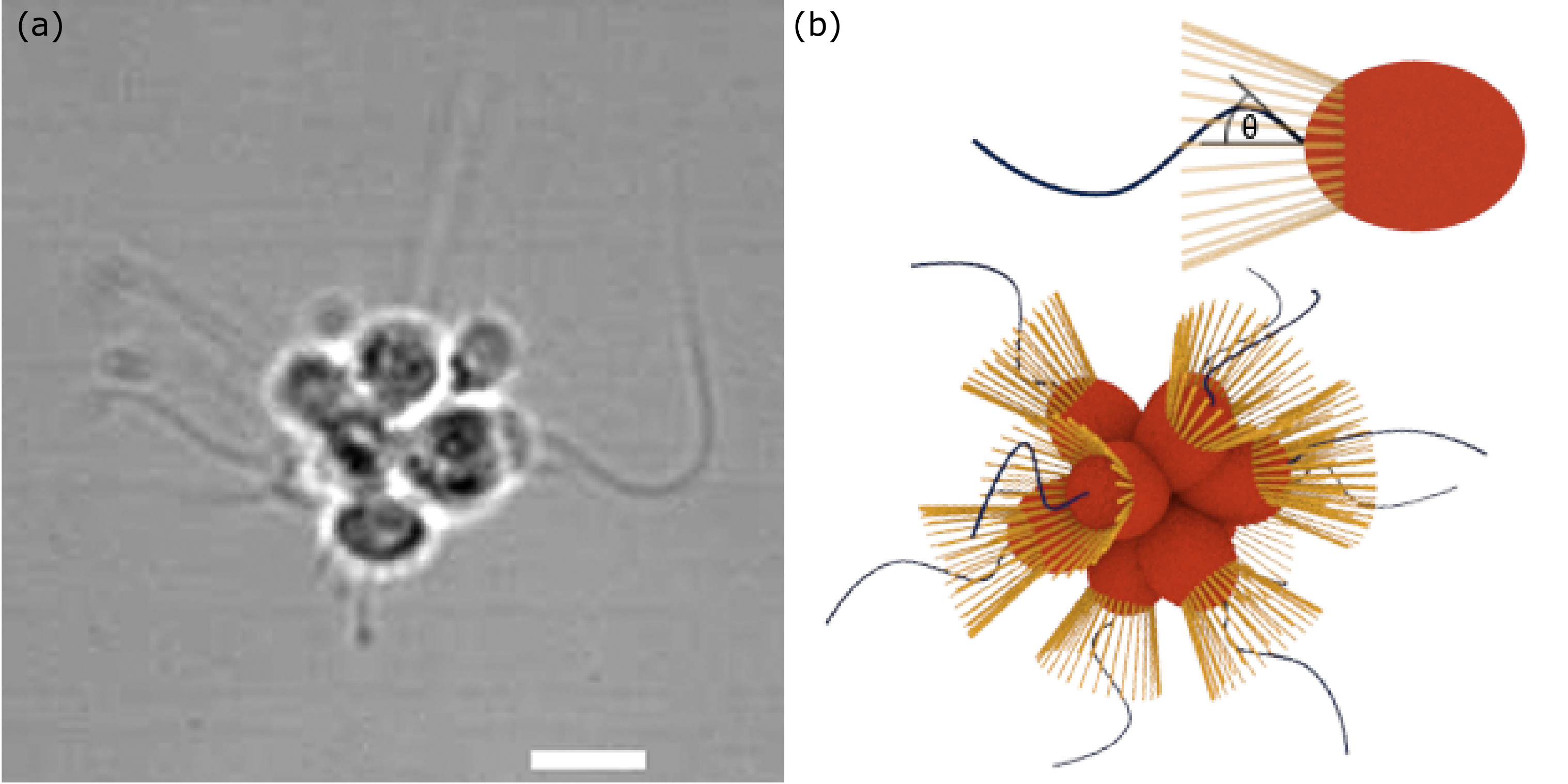}
\caption{(color online) The choanoflagellate {\it S. rosetta}. (a) Bright field image ($5$ $\mu$m scale) and (b) schematics of `slow-swimmer' 
single 
cell, base angle $\theta$, and rosette colony.}
\label{fig:choano}
\end{figure}

Here we study motility in an important example of a `facultative' colonial organism, 
the choanoflagellate \textit{Salpingoeca rosetta} (Fig. \ref{fig:choano}),
which exhibits uni- and multicellular forms with variable cell number.  
Single cells of \textit{S. rosetta}, like other microorganisms, are random walkers (see SI).
We report three main experimental results: (i) individual flagella of the constituent cells beat stochastically, 
(ii) flagella on a given colony display negligible cross-correlation, and (iii) the swimming trajectories of
colonies are stochastic helices.  
These results suggest a hitherto unrecognized 
class of microorganisms, here called \textit{aggregate random walkers} (ARWs): those built by stitching together 
individual random walkers \cite{Poon}. We construct a minimal model to explain this motility.

Choanoflagellates are the closest unicellular relatives of animals \cite{Lang2002}. They filter feed by using their flagellum 
to drive fluid through an eponymous funnel-shaped collar \cite{Pettitt2002}. This beating also confers
motility. 
Fig. \ref{fig:choano} shows a rosette colony of \textit{S. rosetta} which is held together by an extracellular matrix, filopodia, and
intercellular bridges \cite{Dayel2011}. Colonies form by cell division, not aggregation \cite{Fairclough}.  
The evolutionary advantage of the colonial form
is not fully understood, but it is triggered by certain bacteria \cite{Dayel2011,King_eLife}, and 
theory suggests that chain-like 
colonies have enhanced nutrient uptake \cite{Roper2013}.

\begin{figure*}[t]
\centering
\includegraphics{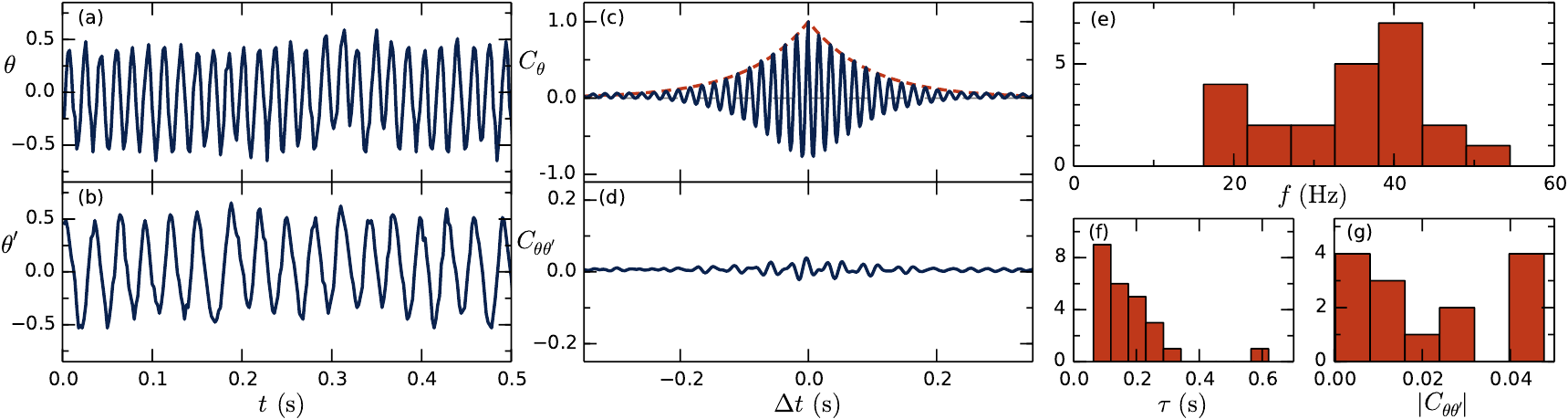}
\caption{(color online) Flagellar beating dynamics. (a,b) Time series of the base angle $\theta(t)$ on two flagella 
within a single colony. (c) Autocorrelation function of $\theta$ for one flagellum, with fit of the envelope to an exponential decay (dashed red). 
(d) Cross correlation of $\theta$ between two flagella on the same colony. (e) Peak frequencies of $n=23$ tracked flagella.
(f) Decay time of autocorrelation in single flagella. 
(g) Magnitude of cross correlations between flagella in same colonies.}
  \label{fig:flags}
\end{figure*}

\textit{S. rosetta}  (obtained from Dr. Barry Leadbeater, University of Birmingham, UK) were cultured in artificial seawater 
($36.5$ g/L Marin Salts (Tropic Marin, Germany)). To provide a food source for prey bacteria, organic 
enrichment ($4$ g/L Proteose Peptone (Sigma-Aldrich, USA), 
$0.8$ g/L Yeast Extract 
(Fluka Biochemika)) was added to the cultures at $15$ $\mu$l/ml. 
Cultures were grown at 22$^{\circ}$C and split every $4-7$ days.
To study the flagella beat 
dynamics, colonies were stuck to poly-L-lysine (0.01\%, Sigma) treated microscope slides and flagella beats imaged at 500 fps 
(Fastcam SA3, 
Photron, USA) in bright field. Image template matching was employed to track the motion of the only slightly 
moving colonies, and 
in the local frame of the organism, the bases of the flagella were tracked by techniques similar to that of the active contour 
model \cite{Kass1988}, yielding as a readout of beating the angle $\theta(t)$ defined in Fig. 1b.

Figs. \ref{fig:flags}a,b show $\theta(t)$ from two flagella on the same colony, and it is clear 
that they have distinct frequencies. In general, the beating frequencies $f$, found by Fourier transforming 
$\theta(t)$, show a surprisingly high variability (Fig. \ref{fig:flags}e). The normalized autocorrelation $C_\theta(\Delta t) = 
\langle \theta(t)\theta(t+\Delta t) \rangle_t/\langle \theta(t)^2 \rangle_t$  for a single flagellum is plotted in Fig. \ref{fig:flags}c.  
Similar to the function discussed \cite{Ma2014, Wan2014} in the context of flagellar beating in {\it Chlamydomonas}, the data
are consistent with $C_\theta=\exp(-|t|/\tau) \cos(2 \pi f \, t)$, the envelope of which is shown in the figure. The decay time 
$\tau$ also shows a very 
high degree of variability (Fig. \ref{fig:flags}f), but all are $<1$ s suggesting high stochasticity. Within colonies, 
the cross correlation between flagella $C_{\theta, \theta'}(\Delta t) = 
\langle \theta(t)\theta'(t+\Delta t) \rangle_t/ \sqrt{ \langle \theta(t)^2 \rangle_t \langle \theta'(t)^2 \rangle_t }$ 
(Fig. \ref{fig:flags}d) is negligible (only a very slight signal can be made out, which we attribute to the overall wiggling 
of the colony -- see
Supplemental Video 1). 
All cross-correlation signals were found to be less than $0.05$ (Fig. \ref{fig:flags}g). The lack of correlation between 
beating flagella in colonies makes \textit{S. rosetta} an ideal model organism for ARWs.

\begin{figure*}[t]
\centering
\includegraphics{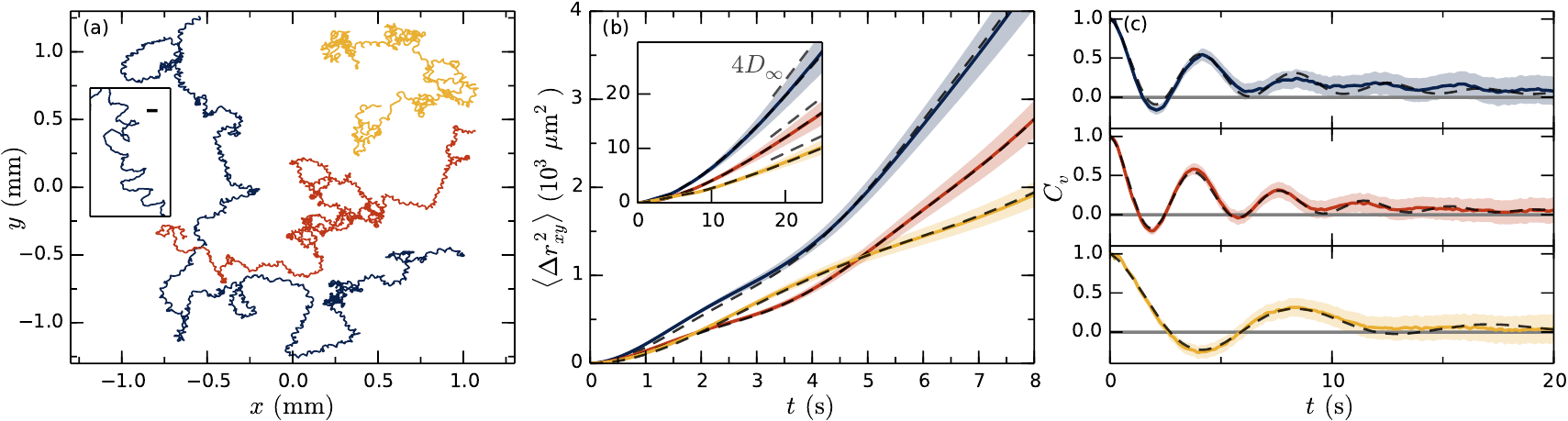}
\caption{(colour online) Random walks. (a) Long tracks of swimming \textit{S. rosetta}. Inset scale bar is $10$ $\mu$m. (b) Projected mean squared displacement
for individual walks (solid, error shaded) and fits of model (dashed). Inset shows a zoom-out with late-time linear behavior $\sim 4 D_\infty t$ (dashed, gray). 
(c) Velocity autocorrelation (solid, error shaded) and model (dashed) with parameters as in (b). 
Fitted parameters (red curve): 
$\omega_0 = 1.63$ s$^{-1}$, $v_p = 8.45$ $\mu$m/s, $v_\omega = 12.7$ $\mu$m/s, $D_r=0.10$ s$^{-1}$.}
\label{fig:randomwalk}
\end{figure*}

In studying the swimming trajectories of \textit{S. rosetta}, ensemble averages taken over many colonies will eliminate 
features related to colony-specific morphology (cell location and flagella orientation). To overcome 
this lack of `ergodicity', we obtained long tracks of $36$ individual colonies. 
In-house software logged and synchronized the position of the $xy$-stage (MS-2000, ASI, USA) 
to a camera (Imaging Source, Germany) filming in bright field at $15$ fps. This enabled tracking of colonies moving in three dimensions at distances 
much longer than the field of view. To track the particles a Gaussian-mixture model \cite{KaewTraKulPong2002} was applied 
to estimate the moving background and subsequently the tracks were manually controlled. Fig. \ref{fig:randomwalk}a shows 
three examples, all $\sim 20$ minutes in length.
On close inspection (inset of Fig. \ref{fig:randomwalk}a) we observe that the
trajectories are noisy helices.  The mean squared displacement 
$\langle \Delta r_{xy}^2 \rangle = \langle [\textbf{x}_{xy}(s+t) - \textbf{x}_{xy}(s)]^2 \rangle_s$ (Fig. \ref{fig:randomwalk}b), 
shows an early time 
ballistic $\sim t^2$ behavior (for $t<1$ s) and late time diffusive $\sim t$ form (inset) similar to that of Janus particles \cite{Howse2007}. 
However, comparing these curves to those of conventional active Brownian particles (see SI), we observe a different intermediate time behavior. These bumps 
(Fig. \ref{fig:randomwalk}b) appear precisely because some of the constituent cells 
may beat off center and induce internal (effective) torques producing stochastic helical
trajectories. 
To highlight the underlying regularity of this
helical swimming, we calculate the velocity autocorrelation $C_v(t) = \langle \textbf{v}_{xy}(s+t) \cdot \textbf{v}_{xy}(s) \rangle_s$ 
(Fig. \ref{fig:randomwalk}c) which oscillates at the frequency of the 
induced rotation and decays on a timescale of several oscillation periods. 

Active random walks have been the attention of much research
\cite{Lovely1974,Lauga2011}, but only recently have rotational torques been incorporated. {\it External} torques 
appear on e.g. magnetotactic bacteria 
in the presence magnetic fields \cite{Blakemore1980} and gyrotactic organisms such as certain algae in gravitational 
fields \cite{Pedley1990} and can be treated analytically \cite{Sandoval2013}. However, the present \textit{internal} 
torques can be treated analytically only in 2D 
\cite{VanTeeffelen2008} and numerical \cite{Wittkowski2012} or approximative \cite{Friedrich2009} methods are needed in 3D. 
Below we develop an approximate 3D theory with the goal of simple analytical functions that can be used to extract physical quantities and interpret the data.

The diffusion of a random walker can be described by the Langevin equation $\d {\bs x}(t) = 
{\bs v}(t) \, \d t + \sqrt{2 D} \, \d {\bs W}(t)$, where $D$ is the translational diffusion constant and ${\bs W}(t)$ a standard 
vector Wiener process with $\langle \d {\bs W}_i(t) \, \d {\bs W}_j(t') \rangle = \delta_{ij} \delta(t-t')$.  
The case ${\bs v}(t) = {\bf 0}$ is a passive particle and leads to the projected mean squared displacement 
$\langle \Delta r_{xy}^2 \rangle \equiv \langle \Delta x^2 
+ \Delta y^2 \rangle = 4 \, D t$. Building an ARW from {\it passive} particles leads to no new behavior, but
motile particles also have a stochastic velocity term. In the simplest case in two dimensions, 
the speed $v$ is constant and 
${\bs v}(t) = v \left( \cos \theta(t), \sin \theta(t) \right)$ evolves stochastically through $\theta(t)$. The choice 
$\d \theta = \sqrt{2 D_r} \, \d W_r(t)$ leads to the conventional result $\langle \Delta r_{xy}^2 \rangle = 
(2v^2/D_r^2) \left(D_r t + e^{-D_r t}-1 \right) + 4 \, Dt$, which behaves ballistically, $\langle \Delta r^2 \rangle \sim t^2$, at early 
times, but diffusively, $\langle \Delta r^2 \rangle \sim t$, at longer times (see SI) with an enhanced diffusion 
constant $D_\infty = D+ v^2/2D_r$, describing well the motion of Janus particles \cite{Howse2007}.
This is a 2D result. Contrary to passive random walkers, active random walkers' effective diffusion constant can vary with dimension.
The corresponding 3D result is $D_\infty = D+ v^2/3D_r$ given a similar definition of 3D rotational diffusion.
Typically, $D \ll v^2/D_r$ and passive diffusion can be ignored.

The Reynolds number for \textit{S. rosetta} is $\text{Re} \sim 10^{-4}$.
At such low Reynolds numbers, inertia is negligible and the fluid dynamics becomes governed by the linear Stokes equation.
Accordingly, self-propelled choanoflagellates are both force- and torque-free.
We assume that \textit{S. rosetta} are sphere-like such that couplings between translations and rotations can be ignored.
Heuristically, the velocity of a colony ${\bs v}(t)$ is approximately a linear sum of the velocities that the constituents would 
have had swimming independently,
${\bs v}(t) \approx \eta \sum {\bs v}_i(t)$, the factor $\eta$ accounting for the change in drag with the radius $a$ of the 
colony, as $\eta \sim a^{-1}$.
If some of the walkers comprising the colony, placed at 
positions $\{ {\bf r}_i \}$, 
beat off center, an angular velocity 
${\bs \omega}(t) \approx \eta_r \sum {\bs r}_i \times {\bs v}_i$ will also be induced, where 
$\eta_r \sim a^{-3}$. 
Since $\{{\bs v}_i\}$ and $\{{\bs r}_i\}$ are given in the local coordinate system of the particle, they must be rotated along 
with the particle. 
For a two dimensional ARW, this motion is described by ${\bs v}(t) = v(t) \left( \cos \theta(t), \sin \theta(t) \right)$, where 
$v(t) = |{\bs v}(t)|$, $\d \theta = \omega(t) \, \d t + \sqrt{2 D_r} \, \d W(t)$, $\omega(t) = \pm |{\bs \omega}(t)|$, and $D_r$ is an effective 
rotational diffusion constant which can be calculated if the individual stochastic processes are prescribed. 
With $v(t)$ constant, constant $\omega(t)$ yields circles in the 
absence of noise. In three dimensions,  such motion leads to helices, making (2D-projected) three dimensional ARWs 
behave very differently from 2D ones and necessitating a full 3D theory.
  
With the goal of a minimal model, we take the swimming speed to be constant and let the direction of the 
velocity evolve with the rotation of the organism according to $\d {\bs v}(t) = \d {\bs \Omega}(t) \times {\bs v}(t)$, 
where $\d {\bs \Omega}(t) = {\bs \omega}(t) \, \d t + \sqrt{2 D_r} \, {\bs \Lambda}(t) \, \d W_r(t)$. 
Here, ${\bs \Lambda}(t)$ is a random unit vector orthogonal to $\v(t)$, uncorrelated in time.
To limit further the number of model 
parameters we assume the magnitude of ${\bs \omega}(t)$ is 
constant, while its direction obeys
$\d {\bs \omega}(t) = \sqrt{2 D_r} \, {\bs \Lambda}(t)  \times {\bs \omega}(t) \, \d W_r(t)$. This last update makes the 
system analytically quite intractable and thus we shall seek an approximate solution.  
As motivation, consider the case $D_r = 0$ with specified initial conditions ${\bs v}(0) = {\bs v}_0$,  ${\bs \omega}(0) = {\bs \omega}_0$. 
This system can be solved exactly to 
yield ${\bs x}(t) = {\bs x}_0 + [( \omega_0 (\w_0 \cdot \v_0)\w_0 t + \w_0 \times (\v_0 \times \w_0) \sin(\omega_0 t) + 
\omega_0 (\w_0 \times  \v_0) ( 1-\cos(\omega_0 t) ) ]/\omega_0^3$, or
\begin{align}
{\bs x}(t) = \int_0^t {\bs v}(t') \, \d t', && {\bs v}(t) = {\bs R} \cdot \begin{pmatrix}
v_\omega \cos \omega_0 t \\
v_\omega \sin  \omega_0 t \\
v_p
\end{pmatrix},
\label{eq:sprial}
\end{align}
where $v_\omega = |{\bs \omega}_0 \times {\bs v}_0|/\omega_0$, $v_p = {\bs \omega}_0 \cdot {\bs v}_0 / \omega_0$, 
$\omega_0 = |{\bs \omega}_0|$, and ${\bs R}$ is some orthogonal matrix. Eq. \eqref{eq:sprial} describes a helix of radius $v_\omega/\omega_0$ and mean speed $v_p$  
(averaged over $2 \pi / \omega_0$). The form of \eqref{eq:sprial} inspires 
an approximative solution in the presence of noise in which
the deterministic helix parameters
define a continuous-time random walk with helix-like steps, 
the matrix ${\bs R}$ becoming a
stochastic matrix process. As an effective description we assume ${\bs R}(t) = {\bs R}_x(\alpha) \cdot {\bs R}_y(\beta) \cdot {\bs R}_z(\gamma)$, 
where the 
matrix factors are rotations around the $x,y,z$ axes.
$\alpha,\beta,\gamma$ are taken independent and identically distributed with $\d \alpha = \sqrt{2 D_r} \, \d W_\alpha(t)$. 
This approximation makes the system much more manageable. 
While the approach breaks $x-y$ symmetry, simulations show it to be 
an overall good approximation for the statistics of interest

In the stationary limit we find (see SI):
\begin{align} \label{eq:autocorr}
\langle {\bs v}_{xy}(\Delta t) \cdot {\bs v}_{xy}(0) \rangle &= 
\frac{e^{-2 D_r | \Delta t | }}{8}  \Big[ 2 v_p^2 \left(1 + 2 e^{D_r | \Delta t | }\right) \\ \nonumber
&+ v_\omega^2 \left(4 + e^{-D_r | \Delta t | }\right) \cos (\omega_0 \Delta t) \Big].
\end{align}
The $xy$-projected mean squared displacement becomes
\begin{align} \label{eq:dr2}
&\langle \Delta r^2_{xy} \rangle = \frac{v_p^2 e^{-D_r t}}{D_r^2} \Big(1+\frac{e^{-D_r t}}{8} \Big) + 4 D_\infty t - a_0 \\ \nonumber
& + v_\omega^2 e^{-2D_r t} \Big(\frac{4D_r^2-\omega_0^2}{(4D_r^2+\omega_0^2)^2}+\frac{(9D_r^2-\omega_0^2)e^{-D_r t}}{4(9 D_r^2+\omega_0^2)^2} \Big) \cos \omega_0 t \\ \nonumber
& -v_\omega^2 e^{-2D_r t} \Big(\frac{4 \omega_0 D_r }{(4D_r^2+\omega_0^2)^2}+\frac{3 \omega_0 D_r e^{-D_r t}}{2(9D_r^2+\omega_0^2)^2} \Big) \sin \omega_0 t,
\end{align}
where the constant $a_0$ enforces $\langle \Delta r^2_{xy} \rangle(t=0) = 0$. 
As $t\to\infty$ we obtain $\langle \Delta r^2_{xy} \rangle = 4 D_\infty t$, where
\begin{equation}
D_\infty = \frac{5v_p^2}{16D_r} + \frac{v_\omega^2 D_r}{8} \left( \frac{1}{6 D_r^2+ \frac{2}{3} \omega_0^2 } + \frac{1}{D_r^2+\frac{1}{4}\omega_0^2} \right).
\label{eq:dinf}
\end{equation}
These results have been verified by simulations using the Euler-Maruyama method. It has previously been shown that reciprocal 
swimming enhances diffusion \cite{Lauga2011}, and the last terms of \eqref{eq:dinf}, which are major 
contributions to the diffusion constant, embody this phenomenon.

Equations \eqref{eq:autocorr} and \eqref{eq:dr2} describe the approximate functions corresponding to the data of 
Figs. \ref{fig:randomwalk}c and \ref{fig:randomwalk}b, respectively. The diffusion constant $D_\infty$ can be 
extracted from the linear late-time behavior of $\langle \Delta r^2_{xy} \rangle$ (dashed gray in inset of Fig.  \ref{fig:randomwalk}b), and can be used in \eqref{eq:dinf} to 
fix one of the model parameters in terms of the others. The remaining three are 
fitted simultaneously to the curves of Fig. \ref{fig:randomwalk}b and \ref{fig:randomwalk}c. 
The experimental data are well-described by the model as shown by the dashed lines in the figures. The relative magnitudes of the extracted velocities, 
$v_p$ and $v_\omega$, reveal how much energy the organisms spent on effective ($v_p$) and circular
($v_\omega$) swimming, for example the blue curve in Fig. \ref{fig:randomwalk} 
has $v_p = 10.6$  $\mu$m/s and 
$v_\omega = 14.2$  $\mu$m/s. While not producing the precise morphologies of the colonies, the fitted velocities combined 
with the extracted frequency $\omega_0$, do constrain the possible configurations. 
Using the fitted velocities and a colony radius $a\sim 5$ $\mu$m, we find an effective translational force of $\sim 1$ pN, and 
using $\omega_0$, an effective torque $\sim 4$ pN$\cdot$ $\!\mu$m: the small residual forces that propel and rotate a
colony are on the order of that of a single cell.

\begin{figure}[t]
\centering
\includegraphics{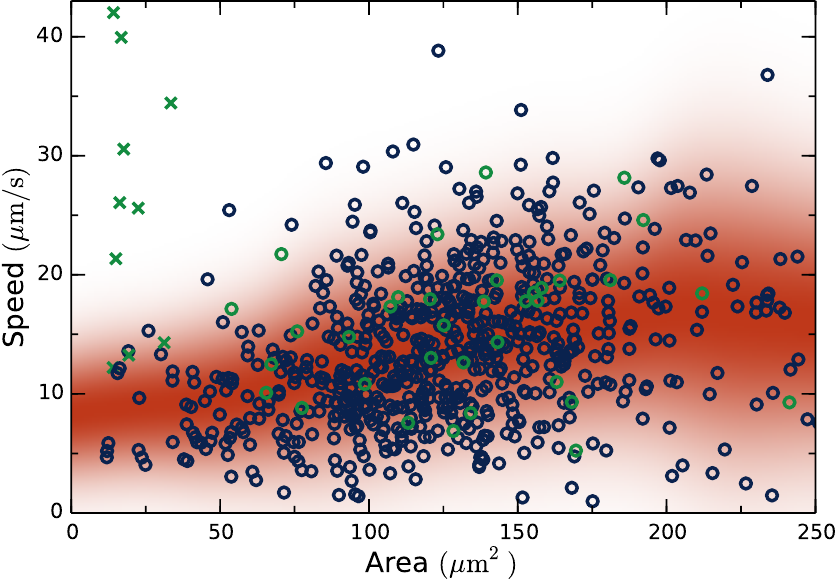}
\caption{(colour online) Speed versus size for $\sim 750$ colonies. Colony size is estimated by median $xy$-projected area. Colonies are blue and green circles, green speed being $\sqrt{v_p^2 + v_\omega^2}$ from model fits to the long tracks. 
Green crosses are single-celled fast swimmers \cite{Dayel2011}. Colored background indicates running mean and standard deviation.}
\label{fig:sizedist}
\end{figure}

Just as flagella beating in \textit{S. rosetta} varies between cells, morphology varies between colonies  
as a result of the cell division process \cite{Fairclough}.  This stochasticity enables two colonies of 
similar size to swim very differently. 
To quantify this, 
we used in-house software to track $\sim 750$ colonies of varying size swimming in quasi-2D between two cover slips, 
and when a colony was in focus the area of an ellipse fitted to its outline served as an estimator of size 
(see Supplemental Video 2). This method, while introducing uncertainty in area, 
enables high throughput. 
To obtain model parameters, long tracks are needed. The parameters for 36 such tracks are given in the SI, and the speed $\sqrt{v_p^2 + v_\omega^2}$ of those tracks is shown in Fig. \ref{fig:sizedist} as green circles.
There is a slow
increase in speed with colony size. This trend can be explained by simple ad-hoc models such 
as random orientation of cells in a sphere-like structure: drag scales linearly with radius $a$ but maximum propulsive force (the case where all propulsive forces point in one direction) 
scales like $a^2$. However, there is an intriguing lack of very 
slow swimmers which would be predicted by such a model. Indeed giving cells an orientation more parallel with its location would 
only yield slower swimming speeds. 
More importantly, Fig. \ref{fig:sizedist} shows just how different colonies of similar size 
are: the stochastic processes underlying colony formation have high variances. 
From fits of the long tracks  this stochasticity seems to apply to all model parameters (SI). This is contrary to e.g. bacterial 
clumps where rotation rate clearly decreases with size \cite{Poon}.
Contrary to the phototactic response of \textit{Chlamydomonas} and \textit{Volvox} in which the time-scale of rotation is matched to inner chemistry \cite{Drescher2010}, 
or the chemotactic response of sperm cells in which curvature and torsion of swimming paths 
are directly manipulated by the single beating flagellum \cite{Friedrich2009}, due to this stochastic morphology of \textit{S. rosetta}, knowledge of 
the overall colony morphology and motion (e.g. $\omega_0$) is arguably not available at the single-cell level, rendering 
`deterministic' chemotactic strategies difficult. Thus one of the most important issues is
the possibility of chemotaxis in aggregate random walks through suitable modulation of the independent constituents \cite{preprint}.

A fundamental operation in the theory of stochastic processes is their summation to yield a single effective process.  The corresponding 
operation for random walkers, `stitching' them 
together, yields ARWs.  As we have shown, there is a crucial complexity for random walkers: the underlying flagellar 
beating can also yield rotations, so the `summation rules' differ. 
Our results suggest that for simple random walkers the ARWs can be described approximately through four numbers: 
$v_\omega$, $v_p$, $\omega_0$, and $D_r$. The question of the correct `summation rules' for general random walkers (e.g. anisotropic, hydrodynamically 
translation-rotation coupled, or non-identical constituents) remains open.
Likewise, the transition, via e.g. self-assembly or flagella growth, from high to low stochasticity in ARWs with non-independent constituents is intriguing.
The present exemplar, \textit{S. rosetta}, is a very good approximation 
to what one might call an ideal biological ARW: independent constituents and a roughly spherical shape.  Its mode of swimming
raises many interesting questions about the evolution of multicellularity and on the nature and origin of noise, 
both internal and environmental.

\begin{acknowledgments}We thank M.E. Cates, E. Lauga, K.C. Leptos and T.J. Pedley for discussions, and an anonymous referee for insightful comments.
Work supported by the EPSRC and St. Johns College (JBK), ERC Advanced Investigator Grant 247333 
and a Wellcome Trust Senior Investigator Award.
\end{acknowledgments}

\bibliographystyle{aipauth4-1}

%\bibliography{Choano}

\begin{thebibliography}{10}

\bibitem{Walther2013}
A. Walther and A.H.E. M{\"u}ller,
Janus particles: synthesis, self-assembly, physical properties, and
  applications, Chem. Rev. {\bf 113}, 5194 (2013).

\bibitem{Howse2007}
J. Howse, R.A.L. Jones, A.J. Ryan, T. Gough, R. Vafabakhsh, and R. Golestanian, Self-motile colloidal particles: from directed propulsion to random
  walk, Phys. Rev. Lett. {\bf 99}, 048102 (2007).
  
\bibitem{Ma2014}
R. Ma, G.S. Klindt, I.H. Riedel-Kruse, F. J{\"u}licher, and B.M. Friedrich,
Active Phase and Amplitude Fluctuations of Flagellar Beating
Phys. Rev. Lett. {\bf 113}, 048101 (2014).

\bibitem{Berg1993}
H.C. Berg, {\it Random walks in biology} (Princeton University Press, Princeton, NJ, 1993).

\bibitem{Polin2009}
M. Polin, I. Tuval, K. Drescher, J.P. Gollub, and R.E. Goldstein,
{\it Chlamydomonas} swims with two “gears” in a eukaryotic version of
  run-and-tumble locomotion, Science {\bf 325}, 487 (2009).

\bibitem{Michelin2010}
S. Michelin and E. Lauga, Efficiency optimization and symmetry-breaking in a model of ciliary
  locomotion, Phys. Fluids {\bf 22}, 111901 (2010).

\bibitem{Brumley2012}
D.R. Brumley, M. Polin, T.J. Pedley, and R.E. Goldstein, 
Hydrodynamic synchronization and metachronal waves on the surface of
  the colonial alga {\it Volvox carteri}, Phys. Rev. Lett. {\bf 109}, 268102 (2012);
R.E. Goldstein, Green algae as model organisms for biological fluid dynamics,
Annu. Rev. Fluid Mech. {\bf 47}, 343 (2015).

\bibitem{Poon} An interesting example of man-made aggregate walkers 
is found in clusters of bacteria induced to form by depletion interactions; J. Schwarz-Linek, C. Valeriani, 
A. Cacciuto, M.E. Cates, D. Marenduzzo, A.N. Morozov, and W.C.K. Poon, Phase separation
and rotor self-assembly in active particle suspensions, Proc. Natl. Acad. Sci. USA {\bf 109}, 4052-4057 (2012).

\bibitem{Lang2002}
B.F Lang, C. O'Kelly, T. Nerad, M.W. Gray, and G. Burger,
The closest unicellular relatives of animals, Curr. Biol. {\bf 12}, 1773 (2002).

\bibitem{Pettitt2002}
M.E. Pettitt and B.A.A. Orme, The hydrodynamics of filter feeding in choanoflagellates,
Eur. J. Protist. {\bf 332}, 313 (2002).

\bibitem{Dayel2011}
M.J. Dayel, R.A. Alegado, S.R. Fairclough, T.C Levin, S.A. Nichols, K. McDonald, and N. King,
Cell differentiation and morphogenesis in the colony-forming
  choanoflagellate {\it Salpingoeca rosetta},
Dev. Biol. {\bf 357}, 73 (2011).

\bibitem{Fairclough}
S.R. Fairclough, M.J. Dayel, and N. King,
Multicellular development in a choanoflagellate,
Curr. Biol. {\bf 20}, 875 (2010).

\bibitem{King_eLife}
R.A. Alegado, L.W. Brown, S. Cao, R.K. Dermenjian, R. Zuzow, S.R. Fairclough, J. Clardy, and N. King,
A bacterial sulfonolipid triggers multicellular development in the closest living relatives of animals,
eLife {\bf 1}, e00013 (2011).

\bibitem{Roper2013}
M. Roper, M.J. Dayel, R.E. Pepper, and M.A.R. Koehl,
Cooperatively generated stresslet flows supply fresh fluid to
  multicellular choanoflagellate colonies, Phys. Rev. Lett. {\bf 110}, 228104 (2013).

\bibitem{Kass1988}
M. Kass, A. Witkin, and D. Terzopoulos, Snakes: active contour models, Int. J. Comp. Vis. {\bf 1}, 321 (1987).

\bibitem{Wan2014}
K.Y. Wan and R.E. Goldstein,
Rhythmicity, recurrence, and recovery of flagellar beating,
Phys. Rev. Lett. {\bf 113}, 238103 (2014).

\bibitem{KaewTraKulPong2002}
P. KaewTraKulPong and R. Bowden, 
An improved adaptive background mixture model for real-time tracking 
with shadow detection, Video-based surveillance systems, 135-144 (2002).

\bibitem{Lovely1974}
P.S. Lovely and F.W Dahlquist, Statistical measures of bacterial motility,
J. Theor. Biol. {\bf 50}, 477 (1974);
E.A. Codling, M.J. Plank, and S. Benhamou,
Random walk models in biology, J. Roy. Soc. Int. {\bf 5}, 813 (2008);
B. ten Hagen, S. van Teeffelen, and H. L\"{o}wen,
Brownian motion of a self-propelled particle, J. Phys. Cond. Matt. {\bf 23}, 194119 (2011).

\bibitem{Lauga2011} E. Lauga, Enhanced diffusion by reciprocal swimming, Phys. Rev. Lett. {\bf 106}, 178101 (2011).

\bibitem{Blakemore1980}
R.P. Blakemore, R.B. Frankel, and A.J. Kalmijn,
South-seeking magnetotactic bacteria in the Southern Hemisphere, Nature {\bf 286}, 384 (1980).

\bibitem{Pedley1990}
T.J. Pedley and J.O. Kessler, A new continuum model for suspensions of gyrotactic
  micro-organisms, J. Fluid Mech. {\bf 212}, 155 (1990).

\bibitem{Sandoval2013}
M. Sandoval, Anisotropic effective diffusion of torqued swimmers, Phys. Rev. E
{\bf 87}, 032708 (2013).

\bibitem{VanTeeffelen2008}
S. van Teeffelen and Hartmut L{\"o}wen, Dynamics of a Brownian circle swimmer,
Phys. Rev. E {\bf 78}, 020101 (2008).

\bibitem{Wittkowski2012}
R. Wittkowski and Hartmut L{\"o}wen, Self-propelled Brownian spinning top: 
dynamics of a biaxial swimmer at low Reynolds numbers, Phys. Rev. E {\bf 85},
021406 (2012).

\bibitem{Friedrich2009} B.M. Friedrich and F. J{\"u}licher, Steering chiral swimmers along noisy helical paths, 
Phys. Rev. Lett. {\bf 103}, 068102 (2009); 
B.M. Friedrich and F. J{\"u}licher, Chemotaxis of sperm cells, Proc. Natl. Acad. Sci. USA {\bf 104}, 13256-13261 (2007).

\bibitem{Drescher2010} K. Yoshimura1 and R. Kamiya, The Sensitivity of Chlamydomonas Photoreceptor is Optimized for the Frequency of Cell Body Rotation, Plant and Cell Physiology \textbf{42}(6) 665-672 (2001). K. Drescher, R.E. Goldstein, I. Tuval, Fidelity of adaptive phototaxis, Proc. Natl. Acad. Sci. USA {\bf 107}, 11171-11176 (2010).

\bibitem{preprint} J.B. Kirkegaard, A. Bouillant, A.O. Marron, K.C. Leptos, 
and R.E. Goldstein, to be published (2015).

\end{thebibliography}

\begin{thebibliography}{10}
\bibitem{PoonSI} J. Schwarz-Linek, C. Valeriani, 
A. Cacciuto, M.E. Cates, D. Marenduzzo, A.N. Morozov, and W.C.K. Poon, Phase separation
and rotor self-assembly in active particle suspensions, Proc. Natl. Acad. Sci. USA {\bf 109}, 4052-4057 (2012).
\end{thebibliography}
%\bibliographystyle{unsrt}

\newpage

\widetext

\centerline{\Large Supplemental Information}
\medskip
\centerline{Motility of Colonial Choanoflagellates and the Statistics of Aggregate Random Walkers}

\author{Julius B. Kirkegaard, Alan O. Marron, Raymond E. Goldstein}

\subsection*{Single cells}

\begin{figure}[htb]
        \centering
          \includegraphics[height=5.5cm]{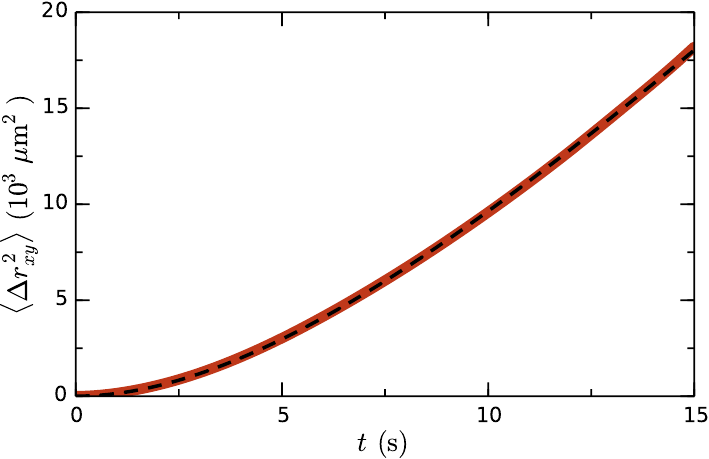}
        \caption{Squared distance moved averaged over 32 \textit{S. rosetta} single cells. Overlayed fit in dashed is to Eq. \eqref{eq:activerandom}. Parameters: $v = 12.3$ $\mu$m/s, $D_r = 0.15$ s$^{-1}$.}
      \label{fig:singlecellplot}
\end{figure}

Slow-swimmer \textit{S. rosetta} unicells, similar in morphology to the individual cells that comprise a colony, clearly exhibit random walk behaviour. Fig. \ref{fig:singlecellplot} shows the average square displacement of 32 S. rosetta slow-swimmers each filmed for $\sim$1.5 minutes. The behaviour is well-described by the equation for conventional active random walkers,
\begin{equation}
\langle \Delta r_{xy}^2 \rangle = (2v^2/D_r^2) \left(D_r t + e^{-D_r t}-1 \right).
\label{eq:activerandom}
\end{equation}

The \textit{active} rotational diffusion constant for both single cells and colonies (main text) are on the order of $0.1$ $s^{-1}$. With a 
beat frequency of $\sim 40$ Hz this corresponds to a $\sim 3^{\circ}$ rotational deviation per beat. The thermal rotational diffusion constant 
$D_r^{\text{thermal}} = k_B T / 8 \pi \mu a^3$ ranges from $0.012$ to $0.0014$ $s^{-1}$ for radii $2.5$ to $5.0$ $\mu$m, at least an order magnitude below the active one.

\subsection*{Derivation of random walker functions}
Since $\alpha$, $\beta$, and $\gamma$ are Markov processes we can write e.g. $P(\alpha(t')) = N(\alpha_0,\penalty 0 2 D_r \,t')$ and $P(\alpha(t) | \alpha(t')) = N(\alpha(t'),\penalty 0 2 D_r \,(t-t'))$ for $t'<t$, where $N(\mu, \sigma^2)$ is the normal distribution. Using $P(\alpha(t),\alpha(t')) \penalty 0 = P(\alpha(t) | \alpha(t'))  \penalty 0 P(\alpha(t'))$ we obtain averages such as
\begin{align}
\langle \cos \alpha(t) \cos \alpha(t') \rangle = \int_{-\infty}^\infty & \d x  \int_{-\infty}^\infty \d y \cos(x) \cos(y) \\ \nonumber
& \times  N_x(\alpha_0, 2 D_r \min(t,t')|) N_y(x, 2 D_r |t-t'|) \\ \nonumber
&= \frac{1}{2} e^{- D_r|t-t'|} \big(1 + \cos(2\alpha_0) e^{-4 D_r\text{min}(t,t')} \, \big),
\end{align}
which in the stationary limit can be used to find the velocity autocorrelations, e.g.
\begin{align}
\langle v_x(t) v_x(s) \rangle & =  \Big\langle \big( v_\omega \cos (\beta(t)) \cos (\gamma(t)) \cos (\omega_0 t) +  v_p \sin(\beta(t))  \\ \nonumber
 & \quad  - v_\omega \cos (\beta(t)) \sin (\gamma(t)) \sin (\omega_0 t)\big) \times \big(v_p \sin(\beta(s))     \\ \nonumber
& \quad  + v_\omega \cos (\beta(s)) \cos (\gamma(s)) \cos (\omega_0 s) - v_\omega \cos (\beta(s)) \sin (\gamma(s)) \sin (\omega_0 s)\big) \Big\rangle \\ \nonumber
& =  \frac{1}{2} \, v_p^2 \, e^{-D_r |t-s|} + \frac{1}{4} \, v_\omega^2  e^{-2D_r |t-s|} \, \cos(\omega_0 (t-s)).
\end{align}
The function only depends on the time difference $t-s$, which is the case for stationary autocorrelations (this is the very definition of a \textit{weakly} stationary process). 
From
\begin{align}
v_y(t) &= -v_p \sin (\alpha(t)) \cos (\beta(t)) + v_w \big( \big[ \sin (\alpha(t)) \sin (\beta(t)) \cos (\gamma(t)) \\ \nonumber
& \quad + \cos(\alpha(t)) \sin(\gamma(t)) \big] \cos(\omega_0 t) + \big[ \cos(\alpha(t)) \cos(\gamma(t)) \\ \nonumber
& \quad - \sin(\alpha(t)) \sin(\beta(t)) \sin(\gamma(t)) \big] \sin(\omega_0 t) \big)
\end{align}
we find in a similar manner
\begin{equation}
\langle v_y(t) v_y(s) \rangle = \frac{1}{4} \, v_p^2 \, e^{-2D_r |t-s|} + \frac{1}{8} \, v_\omega^2 \big( e^{-3D_r |t-s|} + 2 e^{-2D_r |t-s|} \big)  \, \cos(\omega_0 (t-s)).
\end{equation}
Thus
\begin{align} \label{eq:autocorr}
\langle {\bs v}_{xy}(\Delta t) \cdot {\bs v}_{xy}(0) \rangle &= 
\frac{e^{-2 D_r | \Delta t | }}{8}  \Big[ 2 v_p^2 \left(1 + 2 e^{D_r | \Delta t | }\right) + v_\omega^2 \left(4 + e^{-D_r | \Delta t | }\right) \cos (\omega_0 \Delta t) \Big].
\end{align}
The $xy$-projected mean squared displacement is obtained by integrating the autocorrelation twice
\begin{align} \label{eq:dr2}
\langle \Delta r^2_{xy} \rangle &= \int_0^t \int_0^t \langle {\bs v}_{xy}(t') \cdot {\bs v}_{xy}(t'') \rangle \, \d t' \d t'' \\ \nonumber
&=\frac{v_p^2 e^{-D_r t}}{D_r^2} \Big(1+\frac{e^{-D_r t}}{8} \Big) + 4 D_\infty t - a_0 \\ \nonumber
& \quad + v_\omega^2 e^{-2D_r t} \Big(\frac{4D_r^2-\omega_0^2}{(4D_r^2+\omega_0^2)^2}+\frac{9D_r^2-\omega_0^2}{4(9 D_r^2+\omega_0^2)^2} e^{-D_r t} \Big) \cos \omega_0 t \\ \nonumber
& \quad -v_\omega^2 e^{-2D_r t} \Big(\frac{4 \omega_0 D_r }{(4D_r^2+\omega_0^2)^2}+\frac{3 \omega_0 D_r}{2(9D_r^2+\omega_0^2)^2}e^{-D_r t} \Big) \sin \omega_0 t,
\end{align}
where
\begin{equation}
a_0 = \frac{9 v_p^2}{8 D_r^2} + \v_\omega^2 \left( \frac{4 D_r^2 - \omega_0^2}{(4D_r^2 + \omega_0^2)^2} + \frac{9 D_r^2 - \omega_0^2}{4(9D_r^2 + \omega_0^2)^2} \right).
\end{equation}
As $t\to\infty$, $\langle \Delta r^2_{xy} \rangle \sim 4 D_\infty t$, where
\begin{equation}
D_\infty = \lim_{t \rightarrow \infty} \frac{\langle \Delta r^2_{xy} \rangle}{4t}  = \frac{5v_p^2}{16D_r} + \frac{v_\omega^2 D_r}{8} \left( \frac{1}{6 D_r^2+ \frac{2}{3} \omega_0^2 } + \frac{1}{D_r^2+\frac{1}{4}\omega_0^2} \right).
\label{eq:dinf}
\end{equation}
The existence of the above (non-zero) limit confirms the diffusive behaviour.

\subsection*{Comparison of fit parameters}
\begin{figure}[htb]
        \centering
          \includegraphics{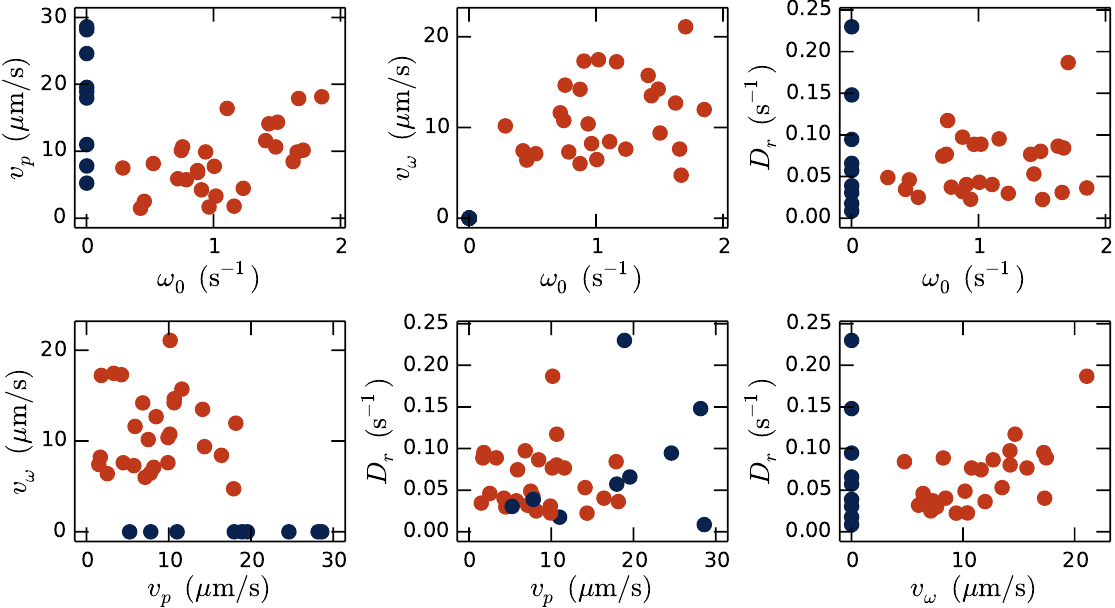}
        \caption{Comparison of fit parameters of 36 tracks. Tracks where $\omega_0, v_\omega$ could be determined in red and tracks where $\omega_0, v_\omega$ are forced to zero in blue.}
      \label{fig:fitparams}
\end{figure}

\begin{figure}[tb]
        \centering
          \includegraphics{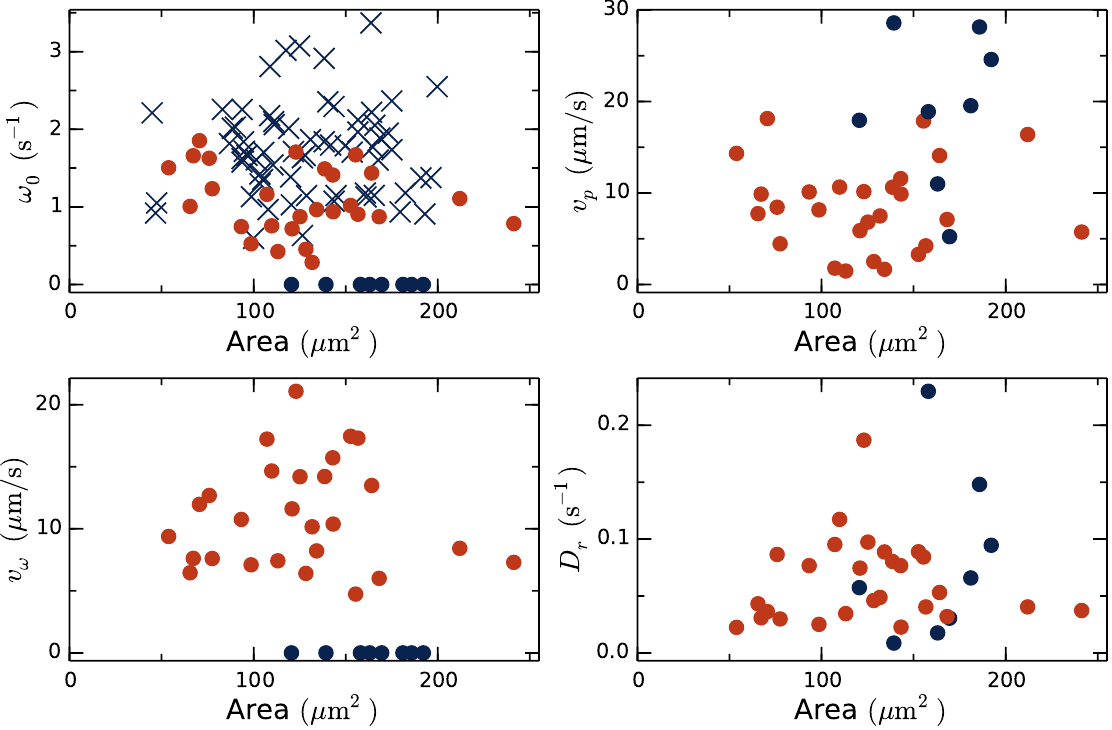}
        \caption{Comparison of fit parameters of 36 tracks to size of colony. Tracks where $\omega_0, v_\omega$ could be determined in red and tracks where $\omega_0, v_\omega$ are forced to zero in blue. $\omega_0$ plot additionally includes estimates from the short track data in blue crosses.}
      \label{fig:sizeparams}
\end{figure}

Fig. \ref{fig:fitparams} shows scatter plots of fit parameters of the model to 36 different \textit{S. rosetta} colonies, indicating the high variances of all parameters. 
We note, however, that the determination of some parameters are difficult in certain regions. 
For instance, $\omega_0$ and $v_\omega$ are hard to determine when either one becomes small, and accordingly we have forced them to zero in these cases and plotted them in blue. 
Naturally, these cases will have a higher $v_p$ as is clear in the two plots in the left part of Fig. \ref{fig:fitparams}.

Applying the same area estimator as in Fig. 4 of the main text, the parameters can also be plotted as a function of size. Just as with swimming speed, Fig. \ref{fig:sizeparams} shows that the model parameters have very high variances and no clear dependency on size. 
For a subset of the short tracks we were able to fit the model well enough to estimate $\omega_0$ and these are shown as blue crosses. However, the short track colonies for which good estimated could be obtained are biased towards high $\omega_0$ (and $v_\omega$). 
Nonetheless, there is no clear tendency for larger colonies to rotate slower as is the case for e.g. bacterial clumps \cite{PoonSI}. For an interesting example of a big fast-spinning colony see end of Supplemental Video 2 in which a colony has formed as a dumb-bell shape.

\pagestyle{plain}

\bibliographystyle{aipauth4-1}

\end{document}